# Investigation of the influence of electrostatic excitation on instabilities and electron transport in E×B plasma configurations

M. Reza*[1], F. Faraji*, A. Knoll*, B. Rose**

*Department of Aeronautics, Imperial College London, London, United Kingdom

**Orbit Fab, Harwell, England, United Kingdom

**Abstract**: Partially magnetized plasmas in E×B configurations – where the electric and magnetic fields are mutually perpendicular – exhibit a cross-field transport behavior, which is widely believed to be dominantly governed by complex instability-driven mechanisms. This phenomenon plays a crucial role in a variety of plasma technologies, including Hall thrusters, where azimuthal instabilities significantly influence electron confinement and, hence, device performance. While the impact of prominent plasma instabilities, such as the electron cyclotron drift instability (ECDI) and the modified two-stream instability (MTSI) on cross-field transport of electron species is well recognized and widely studied, strategies for actively manipulating these dynamics remain underexplored. In this study, we investigate the effect of targeted wave excitation on instability evolution and electron transport using one- and two-dimensional particle-in-cell simulations of representative plasma discharge configurations. A time-varying electric field is applied axially to modulate the spectral energy distribution of the instabilities across a range of forcing frequencies and amplitudes. Our results reveal that the so-called "unsteady forcing" can both suppress and amplify instability modes depending on excitation parameters. In particular, across both 1D and 2D simulation configurations, forcing near 40 MHz effectively reduces ECDI amplitude and decreases axial electron transport by about 30%, while high-frequency excitation near the electron cyclotron frequency induces spectral broadening, inverse energy cascades, and enhanced transport. These findings point to the role of nonlinear frequency locking and energy pathway disruption as mechanisms for modifying instability-driven transport. Our results offer insights into potential pathways to enhance plasma confinement and control in next-generation E×B devices.

**Section 1: Introduction**

Partially magnetized plasmas subject to mutually perpendicular electric and magnetic fields, commonly referred to as E×B or cross-field plasmas, have found widespread applications across various industries over recent decades. These applications include magnetrons for plasma-assisted manufacturing, Hall thrusters for spacecraft propulsion, and plasma sources such as Penning discharges. E×B plasmas are characterized by phenomena that span a broad range of spatial and temporal scales [1][2]. The global dynamics of plasma discharge – and consequently the performance of E×B devices – are driven by the intricate interactions among the underlying plasma processes, including instabilities and turbulent mechanisms, which affect the plasma species via complex and highly coupled dynamics. This coupling between instabilities and turbulent processes not only makes it difficult to predict plasma transport patterns but also poses significant challenges for controlling and optimizing the behavior of plasma in the E×B configurations. As a result, achieving effective controlling mechanisms over naturally occurring plasma instabilities represents a crucial step toward improving the efficiency and stability of cross-field plasma devices.

The operation principle of many E×B technologies largely relies on effective confinement of electrons to limit their transport across the magnetic field lines. This is essential to maintain efficiency and performance, particularly in devices reliant on propellant ionization via direct electron impact, such as Hall thrusters. It is now widely accepted that a major contributor to electron transport in cross-field plasma discharges, such as in Hall thrusters, is the presence of instabilities along the azimuthal (circumferential) direction, which is aligned with the electrons' E×B drift [3]-[6]. The resulting electron transport has been shown to arise from correlations between oscillations in electron number density and the azimuthal electric field [7]. The resulting effect is interpreted as an instability-enhanced "friction" force between electrons and ions [8][7][9]. Given the substantial role that azimuthal instabilities play in driving cross-magnetic-field electron transport, it is anticipated that suppressing or mitigating the amplitudes of these instabilities may result in reducing the significance of electron transport induced by these mechanisms.

Modulating anode voltage is a known technique for controlling the low frequency macroscopic unstable behaviors exhibited by Hall thrusters. Simmonds et al. [11] investigated methods for using a modulated electric field to manipulate the breathing mode of a Hall thruster, demonstrating minor performance improvements. Romadanov et al. [11] demonstrated active control of the spoke mode using a similar approach. Both techniques applied a modulated electric field to the plasma at a relatively low driving frequency, on the order of kHz for the characteristic frequency of the breathing mode and the spoke mode, respectively. In another example, Tejeda et al. [12] performed a numerical experiment at a much higher frequency of excitation, on the order of the electron cyclotron frequency, to assess possible improvements in efficiency, thrust and specific impulse of a Hall thruster. An analogous approach has been also adopted to suppress drift wave instabilities in Q-machines, such as the experimental and theoretical works of Weiss and Morrone [13], as well as Y. Nishida et al. [14].

---

[1] **Corresponding Author** (m.reza20@imperial.ac.uk)

Studies investigating specifically the effects of varying azimuthal magnetic field strength and neutral density on electron transport have been carried out by Bak et al [15][16]. It was shown through experiments and simulation that azimuthal non-uniformities in the neutral propellant density induced axial-azimuthal electric fields which contributed to the electron transport [15]. Additionally, a Hall thruster with a non-uniform azimuthal magnetic field was tested to provide insights into the resulting electron mobility [16]. An alternative approach has also been proposed by Kapulkin and Prisnyakov [17], who investigated theoretical methods for suppressing the electron drift instability by making special grooves on the channel walls of a HET in order to affect the electron surface collision frequency. Most recently, Rose and Knoll demonstrated methods to increase or decrease the electron transport using a modulated axial field in a 1D simulation by directly targeting excitation at the fundamental frequency of the Electron Cyclotron Drift Instability (ECDI) [18].

In this study, we investigate how externally applied electrostatic excitation influences plasma response in both one- and two-dimensional E×B discharge configurations. We aim to assess how the applied modulations at various frequencies and with different amplitudes may affect instabilities spectra and the consequent wave-induced transport, identifying the modulation conditions corresponding to reduction and increase in the cross-field electron mobility. This enables insights into potential mechanisms for instability control and transport regulation in E×B devices.

**Section 2: Theoretical background**

**2.1: Overview of the characteristics of dominant azimuthal instabilities**

A key azimuthal instability in E×B discharges like in a Hall thruster is the ECDI – a high-frequency (1–10 MHz) kinetic instability with wavelengths on the scale of the electron Larmor radius. This instability propagates mainly along the azimuthal (E×B) direction. In typical conditions of E×B discharges such as Hall thrusters, the phase velocity of the waves is lower than the azimuthal drift of the electrons, allowing the waves to grow by extracting energy from the drifting electrons. The ECDI excites as a result of coupling between Doppler-shifted electrostatic Bernstein modes and the ion acoustic waves [6]. Extensive research has been conducted on ECDI over the years through theoretical models [9][19], numerical simulations [20]-[27], and experimental studies [28][29]. The significance of the ECDI roots in the multitude of effects it has on plasma behavior, in particular increased electron cross-field transport [4][8][22][30] and significant plasma heating [7][24][25][31].

The 2D approximation of the ECDI's dispersion relation, perpendicular to the applied magnetic field (in the axial-azimuthal plane), reveals that the ECDI unstable modes develop close to the resonance condition specified by Eq. 1 [8][19]

$$k_z = n \frac{\omega_{ce}}{V_{de}} = n \frac{e}{m_e} \frac{B^2}{E} \quad ; \quad n = 1, 2, ..., \qquad \text{(Eq. 1)}$$

where, $k_z$ represents the wave's azimuthal wavenumber, $\omega_{ce}$ is electron cyclotron frequency, $V_{de}$ is electrons' azimuthal drift velocity, and $E$ and $B$ are the magnitudes of the axial electric field and radial magnetic field, respectively.

Another influential instability in E×B plasmas is the modified two-stream instability (MTSI). McBride's theoretical analysis [32] demonstrated that this mode behaves as a fluid-like instability, unaffected by the electron-to-ion temperature ratio ($T_e/T_i$). Numerical studies by Janhunen et al. [33] and Taccogna et al. [34] identified this instability as a relatively long-wavelength mode with a non-zero wavenumber in the radial direction. The presence of this mode in the systems studied contributed to an enhanced inverse energy cascade phenomenon.

Boeuf and Smolyakov in Ref. [2] presented the MTSI dispersion relation by adapting the kinetic ECDI dispersion relation, specifically for the case of cold electrons ($T_e \to 0$) and including a finite wavenumber component along the magnetic field direction. Petronio et al. [35][36] obtained a simplified 2D fluid dispersion relation for the MTSI in the radial-azimuthal plane, assuming isothermal electrons with isotropic temperature and treating ions as cold and unmagnetized. They then derived a simple relationship between the radial and the azimuthal wavevector components of the MTSI's fastest growing mode as in Eq. 2 [35]

$$k_z = \sqrt{\frac{e}{m_e} \frac{B^2}{E} k_x},$$  (Eq. 2)

where, $k_z$ and $k_x$ denote the wave's azimuthal and radial wavenumber components, respectively.

**2.2: Wave excitation in plasmas and energy transfer mechanisms**

The method of wave excitation investigated in this study is to apply a sinusoidal time varying electric field in a direction perpendicular to both an external steady magnetic field and the E×B drift of the electrons (described in detail in Section 3). The frequency of the imposed electric field oscillation will interact with the natural dynamic modes of the system. If the system was purely linear, then the frequency of excitation would match exactly the corresponding response of the system, which could be measured, for instance, in properties such as the time variation of the plasma density. However, the plasma dynamics are nonlinear, and the response of the system is therefore much more complex.

This study focused on two particular behaviors: frequency locking and unsteady forcing of energy pathways between natural oscillatory modes of the plasma. Frequency locking can be described as the tendency of naturally occurring oscillations in the plasma system which are close to the excitation frequency to become locked to the driving frequency. For instance, if the driving frequency is close to, but not exactly, the same as the electron cyclotron frequency, then the cyclotron frequency will be slightly altered to lock into the driving frequency exactly. It should be noted that such behavior cannot occur with a linear description of the plasma dynamics and rely on the nonlinearity of the system to accommodate such a subtle change in the resonance conditions.

Unsteady forcing of the energy pathways can be described as a promotion or disruption to natural coupling between oscillatory modes of the plasma. An example can be the transfer of energy between high frequency short wavelength modes to long wavelength low frequency modes known as inverse energy cascades [37]. For example, if the frequency of a low energy mode is drawn away from a harmonic resonance with a high frequency oscillation through frequency locking, then the energy transfer between the two modes is disrupted. Conversely, if a low frequency mode is aligned in resonance with a high frequency oscillation, then a new energy pathway is created. Since the energy pathways of the system are changed, the macroscopic impact of the driving oscillation is not limited by the energy of the driving oscillation itself. An exciting possibility is that large scale variations of the quasi-steady behavior of the system could result even through low energy forcing of the system at specific frequencies.

**Section 3: Setup of the simulations performed**

**3.1. Simulations' setup and conditions in the 1D azimuthal configuration**

The configuration of this case corresponds to the azimuthal coordinate of a Hall-thruster-representative discharge with the settings that closely follow those reported in Ref. [8]. The radial, axial and azimuthal directions of the coordinate system for this simulation setup are denoted by *x*, *y*, and *z*, respectively. The domain's size is 0.5 cm, which is discretized using $N_i = 100$ computational nodes. A uniformly distributed and stationary electric field ($E_{y,0}$) with the magnitude of $20\ kVm^{-1}$ and a constant radial magnetic ($B_x$) with an intensity of 20 mT are applied. The simulation is initialized by loading electron and ion particles with a uniform density of $1 \times 10^{17}\ m^{-3}$ inside the domain. The initially loaded electrons and ions are sampled from Maxwellian distributions at the temperatures of 2 eV and 0.1 eV, respectively. The number of macroparticles per cell for either electron or ion species is 200. The time step is $5 \times 10^{-12} s$, and the plasma properties are averaged and recorded every 20 timesteps. Noting the fact that the dominant physics of interest along the azimuthal coordinates, namely the excitation, growth and saturation of the involved azimuthal instability, is primarily collisionless in nature, the collision are neglected in the simulations.

To represent the axial convection of the instability waves and, hence, to limit the growth of instability waves, a fictitious axial extent with the length of 1 cm is considered, which is consistent with the adopted value in Ref. [8]. The electrons and ions that leave the axial boundary on one side are resampled from their initial distribution and reinjected into the domain from the opposite axial boundary. Along the azimuthal direction, a periodic boundary condition is applied on the particles, while along the radial coordinate particles are allowed to move freely. Note that, even though a finite axial length is assumed, the simulation remains a 1D problem. This is due to the fact that Poisson's equation is solely solved along the azimuthal coordinate. To guarantee an azimuthally periodic solution of the plasma potential, zero-value Dirichlet conditions are imposed at both azimuthal boundaries of the domain.

The described setup represents the baseline simulation. The total simulated time is 20 $\mu s$. To study the response of plasma to unsteady forcing, a temporally oscillatory electric field is applied on top of the constant axial electric field ($E_{y,0}$). Therefore, in these "forced" simulations, the axial electric field at each timestep is $E_y = E_{y,0} + A_F \sin(2\pi\omega_F t)$, where $\omega_F$ is the forcing frequency and $A_F$ is the forcing amplitude. The forced simulations are initialized from the quasi-steady state of the baseline case at the end of 10 $\mu s$. The forcing amplitude is chosen to be 50 % of the stationary axial electric field ($A_F = 0.5 E_{y,0} = 10\ kVm^{-1}$). Forced simulations are performed for various forcing frequencies across a range of $1 - 800$ MHz. For each forcing frequency, simulations are repeated eight times (each for 20 $\mu s$ duration) to provide a statistically representative result.

The IPPL[1]-1D PIC code is used for simulations. The code has been verified and benchmarked in Refs. [38][39].

### 3.2. Simulations' setup and conditions in the 2D radial-azimuthal configuration

The simulation configuration resembles the radial-azimuthal cross-section of a typical Hall thruster with the numerical and physical setup being similar to the conditions of the community benchmark problem reported in Ref. [40]. The computational domain represents a 2D Cartesian plane with equal lengths of 1.28 cm along both simulation directions ($L_x = L_z = 1.28$ cm). In terms of axis notation, $x$, $y$, and $z$ denote, respectively, the radial, axial, and azimuthal directions. The stationary applied axial electric field ($E_{y,0}$) has a magnitude of $10\ kVm^{-1}$ and the external radial magnetic field's ($B_x$) intensity is 20 $mT$. The schematic of the simulation domain and the directions of the imposed electric and magnetic fields relative to the coordinate system are illustrated in Figure 1.

The computational cell size is considered to be 50 $\mu m$, which corresponds to 256 nodes along both simulation dimensions. The time step is $1.5 \times 10^{-11}$ s. The azimuthal electric field signal, which is used to analyze the instabilities' spectra are averaged and reported every 20 timesteps.

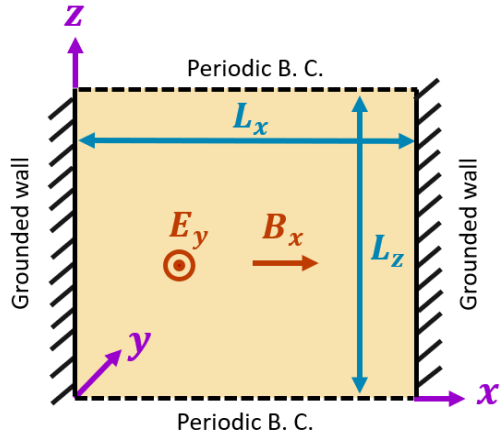

Figure 1: Schematics of simulation's computational domain, coordinate system, and the applied stationary electric and magnetic fields

Initially, electrons and ions are sampled from Maxwellian distribution functions at the temperatures of 10 eV and 0.5 eV, respectively, and are loaded uniformly in the domain with a density of $1.5 \times 10^{16}\ m^{-3}$. The initial number of macroparticles per cell is 100. The simulations are collisionless, and the establishment of a steady state in the simulation setup is achieved through introducing a particle injection source. The injection source is azimuthally uniform and follows a cosine profile along the radial direction, spanning from $x = 0.09$ cm to $x = 1.19$ cm, with the peak value of $8.9 \times 10^{22}\ m^{-3}s^{-1}$. The electron-ion pairs are sampled from Maxwellian distribution functions at the initial temperatures of the respective species and injected into the domain following the distribution prescribed by the injection source.

Regarding particle boundary conditions, any particle reaching the wall boundaries is eliminated, and no secondary electron emission is considered. To reflect periodicity condition along the azimuthal coordinate, particles crossing the azimuthal boundaries are reinjected from the opposing boundary while retaining their velocity and radial position. Similar to the 1D azimuthal case, as the simulations do not resolve the axial direction, a finite artificial extent of 1 cm is assumed along the $y$ direction on both sides of the radial-azimuthal simulation plane [40]. Particles reaching the axial boundaries are resampled from the initial Maxwellian distributions and reloaded onto the simulation plane at the same radial and azimuthal positions.

---

[1] Imperial Plasma Propulsion Laboratory

As for the boundary condition for the electric potential, a zero-volt Dirichlet condition mimics the grounded walls along the radial coordinate. A periodic condition is applied along the azimuthal direction.

The simulations are performed using the reduced-order IPPL-Q2D code with a domain decomposition [41] corresponding to 50 regions along both the radial and azimuthal directions. At this approximation level, the Q2D simulations are demonstrated to reproduce faithfully the results from a traditional full-2D PIC code [42].

The baseline simulation features only the stationary axial electric field ($E_y = E_{y,0}$). Also, the total simulated time in the baseline condition is 30 $\mu s$. In the forced simulations, similar to the 1D setup in section 3.1, the imposed axial electric field involves an oscillatory component. Specifically, the electric field is represented as $E_y = E_{y,0} + A_F \sin(2\pi \omega_F t)$, with $\omega_F$ and $A_F$ denoting the forcing frequency and the forcing amplitude, respectively. The state of the plasma at end of the baseline simulation at 30 $\mu s$ is used as the initial condition in the forced simulations. We assessed three different forcing frequencies including ion plasma frequency ($\omega_{pi}$ = 5.8 MHz), electron cyclotron frequency ($\omega_{ce}$ = 560 MHz) and a mid-range frequency in between the previous two (40 MHz). Each frequency is assessed at four different forcing amplitudes of 5 %, 10 %, 25 % and 50 % of the stationary axial electric field's magnitude. We performed three 15 $\mu$-long simulation repetitions (from 30 $\mu s$ to 45 $\mu s$) for every forcing condition to ensure statistical significance of the observed behaviors.

**Section 4: Results**

In this section, we discuss the impact of the wave excitation on the frequency and wavenumber spectra of the azimuthal instabilities and, in turn, on the axial electron mobility induced by the excited instabilities. We first present the outcomes for the 1D azimuthal case and then proceed to the results in the radial-azimuthal setting.

The spectra of instabilities in the 2D simulations exhibit notable differences from those observed in the 1D simulation. Consequently, the comparison between the 1D and 2D results enables us to evaluate how the effect of the unsteady forcing on the plasma might be different in the presence of the radial physics.

**4.1. 1D azimuthal configuration**

Figure 2 provides a comparison of the frequency spectra of the instabilities in the absence (baseline simulation) and in the presence of the unsteady forcing with various frequencies. The frequency spectrum is derived using temporal fast Fourier transform (FFT) analysis of the azimuthal electric field ($E_z$) signal over a duration of $4 - 6$ $\mu s$ at a certain azimuthal location. This duration refers to the time interval after arriving at quasi-steady state at 10 $\mu s$, hence, the absolute time of 14-16 $\mu s$. The same is true wherever referring to the time interval in this section. The resulting spectrum in each scenario is the average of the FFTs of $E_z$ across all azimuthal positions and over eight simulation repetitions.

In addition, the azimuthal wavenumber ($k_z$) spectra of the instabilities in the forced simulations are compared against the baseline $k_z$ spectrum in Figure 3. The $k_z$ spectrum in each case is obtained by calculating the spatial FFT of the $E_z$ signal across the entire domain at a specific time instance between $4 - 6$ $\mu s$. The plots then represent the average of the spatial FFTs over the full $4 - 6$ $\mu s$ duration and across eight simulation repetitions.

In the baseline 1D setup, the ECDI establishes within the simulation as the dominant instability with the frequency of 4.5 MHz and azimuthal wavenumber of 618.8 $m^{-1}$ (wavelength of about 1.6 mm). The plots in Figure 2 and Figure 3 show that the application of the unsteady forcing changes significantly the spectral amplitude of the instabilities. The specific impact varies depending on the frequency of the forcing ($\omega_F$). At very low frequency ($\omega_F$ = 1 MHz), the ECDI's peak in the frequency spectrum is flattened, and its amplitude is nearly halved compared to its original value in the baseline. The energy from the ECDI's waves seems to have transferred to the fluctuations of mid-range frequencies between $40 - 250$ MHz. Slightly increasing the forcing frequency (cases with $\omega_F$ = 4.5 and 5.8 MHz) results in a shift of the peak of the frequency spectra towards the frequency of the ion acoustic instability ($\omega_{IA} \approx$ 3.35 MHz, as calculated from $\omega_{IA} \approx \frac{k_z \lambda_D \omega_{pi}}{\sqrt{1+k^2\lambda_D^2}}$). The peak wavenumber also migrates towards $k_z = 412$ $m^{-1}$ corresponding to the longer wavelength of about 2.4 mm. In these cases, the spectral amplitude of an extended range of frequencies across $15 - 250$ MHz is increased.

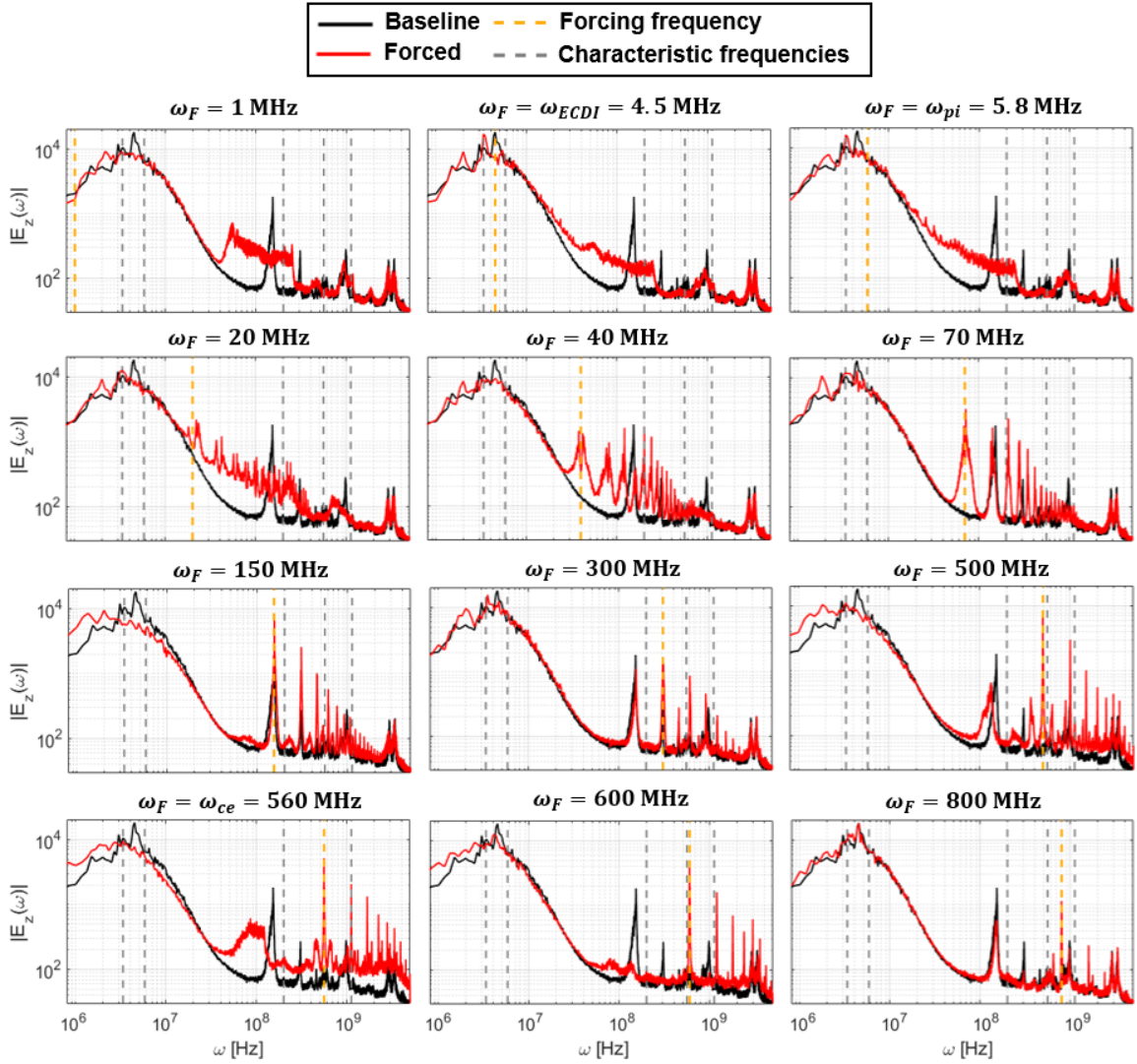

Figure 2: Variation of the frequency spectrum relative to the baseline case from 1D azimuthal simulations with various forcing frequencies. In each case, the spectrum represents the average of the temporal FFT of azimuthal electric field ($E_z$) signal over all azimuthal positions and over 8 simulation repetitions. The characteristic frequencies displayed in grey dashed lines from left to right correspond to theoretical ion acoustic frequency ($\omega_{IA}$), ion plasma frequency ($\omega_{pi}$), Hall circulation frequency ($\omega_E$) and first and second harmonics of electron cyclotron frequency ($\omega_{ce}, 2\omega_{ce}$).

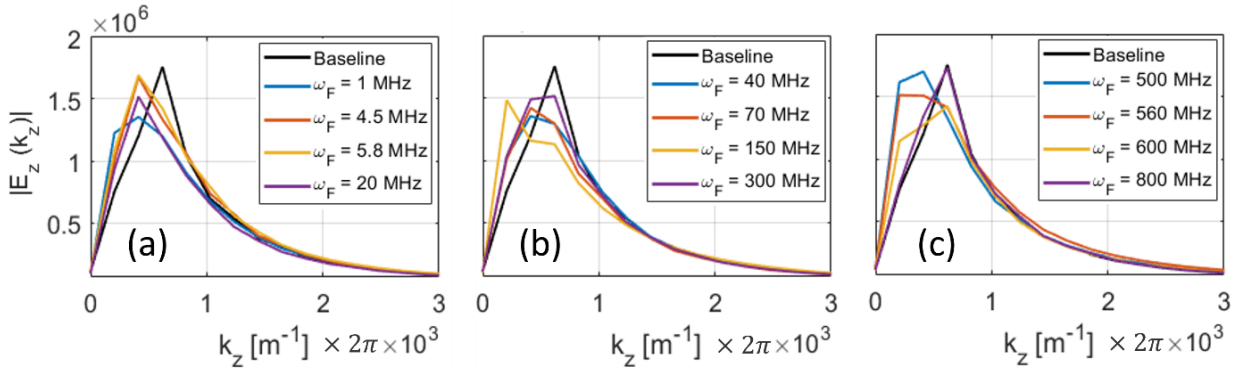

Figure 3: Variation of the azimuthal wavenumber ($k_z$) spectrum relative to the baseline case from 1D azimuthal simulations with various forcing frequencies. In each case, the spectrum represents the temporally averaged (over 4 to 6 $\mu s$) spatial FFT of azimuthal electric field ($E_z$) signal averaged over 8 simulation repetitions. Plots (a)-(c) denote varying frequency ranges.

Forcing with the frequency range between $20 - 70$ MHz excites instabilities with frequencies in the range of $\geq \omega_F \leq \omega_{ce} = 560$ MHz), with the distribution of energy being mostly concentrated on the discrete peaks. At forcing frequencies of $\omega_F = 150$ and $500$ MHz, a significant reduction in the amplitude of ECDI is evident. Meanwhile, the oscillations within the low-frequency range of the spectra ($\omega < \omega_{AI}$) and longer wavelengths are amplified. In these cases, the amplitudes of the high frequency range of the spectra experience a slight increase. Interestingly, these variations are not observed in the spectral amplitudes in the case with $\omega_F = 300$ MHz.

Excitation at higher frequencies near the electron cyclotron frequency ($\omega_F = \omega_{ce} = 560$ MHz) intensifies the high frequency range of the spectrum while having similar effect to cases with $\omega_F = 150$ and 500 MHz on the frequency range below 30 MHz. These variations in the spectral amplitude of the instabilities appear to diminish at higher forcing frequencies such as $\omega_F = 600$ MHz, and nearly disappear when the forcing is applied at $\omega_F = 800$ MHz.

As an interesting observation, in all cases with forcing frequency $\omega_F \leq 70$ MHz (and at $\omega_F = 300$ MHz), a peak between $1.8 - 2.2$ MHz appears in their frequency spectrum. Whereas in cases with $150$ MHz $\leq \omega_F \leq 600$ MHz, the entire range of low frequency spectrum below ion acoustic frequency is enhanced.

Looking at Figure 3, it is evident that unsteady forcing in most cases tend to shift the $k_z$ spectra towards smaller wavenumbers, effectively increasing the wavelength of the dominant instabilities.

The variations in the spectral amplitude of the azimuthal instabilities will inherently translate into change in the electrons' cross-field (axial) transport caused by these fluctuations. To quantify the change in the electrons' transport, the axial current density ($J$) and electrons' axial mobility ($\mu$) in the presence of the unsteady forcing at various frequencies are compared against their values in the baseline simulation in Figure 4. These quantities are obtained from the spatiotemporal signals of the electron number density ($n_e$) and the electrons' axial drift velocity ($V_{e,y}$) from the simulations according to the following relations

$$J = \frac{1}{N}\sum_{k=1}^{N} \frac{e}{L_z(t_2-t_1)} \int_{t_1}^{t_2} \int_0^{L_z} n_e^k(z,t)\, u_e^k(z,t)\,, \qquad (\text{Eq. 3})$$

$$\mu = \frac{1}{N}\sum_{k=1}^{N} \frac{1}{L_z(t_2-t_1)E_{y,0}} \int_{t_1}^{t_2} \int_0^{L_z} V_{e,y}^k(z,t)\,, \qquad (\text{Eq. 4})$$

where, $t_1 = 4\ \mu s$ (14 $\mu s$ in absolute time) and $t_2 = 10\ \mu s$ (20 $\mu s$ in absolute time), and $N$ is the total number of repetition instances for each forcing scenario which in this case is 8. Moreover, to determine the statistical significance of the observed variations in the electrons' axial current and mobility, the first standard deviation among the eight-simulation set in each forcing case is indicated in the plots of Figure 4 as well. Also, the normalized variation percentages of these quantities with respect to their baseline values, which are shown in Figure 4, are calculated according to the following

$$|\overline{\Delta J}| = \frac{1}{N}\sum_{k=1}^{N}\left|\frac{J_F^k - J_B^k}{J_B^k}\right|, \qquad |\overline{\Delta \mu}| = \frac{1}{N}\sum_{k=1}^{N}\left|\frac{\mu_F^k - \mu_B^k}{\mu_B^k}\right|. \qquad (\text{Eq. 5})$$

In the above relations, the subscript "$F$" and "$B$" denotes the quantities in the forced and baseline simulations, respectively.

The plots in Figure 4 show that, overall, forcing at frequencies below 150 MHz leads to a reduction in the electrons' axial current density and electrons' axial mobility by an average of about $15 - 30\%$ relative to the baseline case. It is noticed that the maximum reduction occurs at $\omega_F = 40$ MHz, which corresponds to about 29%. The decrease in the electrons' transport is consistent with the observation that, in these cases, the amplitude of the ECDI's instability as a major contributor to the axial transport of electrons is diminished.

Excitation at frequencies equal to and above 150 MHz either enhances electron axial transport or results in no significant change relative to the baseline. It is noted that excitation at this frequency range substantially amplifies the waves close to electron cyclotron frequency, which can resonate with the electrons' cyclotron motion and increases the electrons' Larmor radius, thereby weakening magnetic confinement. This can partially be responsible for the observed enhanced transport in this range of excitation frequency. The increase in electrons' current in cases of $\omega_F = 150, 500$ and 560 MHz is correlated with the amplification of the longer wavelength modes, which can contribute to the enhanced transport. In these cases, despite the suppression of the ECDI's peak, the combined current carried through the longer wavelength instabilities and the increased transport due to reduced magnetization of electrons compensate for, and even exceed, the current that would have otherwise been induced by the ECDI waves. In particular, forcing frequencies of 150 MHz and 500 MHz increase the electrons' current by about 26% and 21%, respectively, while increasing the electrons' axial mobility by about 13% in both cases.

This is in contrast to the excitation frequencies below 150 MHz, which results in changing both $J$ and $\mu$ by almost the same percentage relative to their baseline values. The observed difference in the behavior of $J$ and $\mu$ is related to the different degree of correlations between the electron number density and the electron axial velocity signals at various excitation frequencies.

When exciting at the electron cyclotron frequency $\omega_F = \omega_{ce} = 560$ MHz, the electrons' axial currents and mobility increase by approximately 12% and 8%, respectively.

The variations in both the electrons' current and mobility for the remaining forcing frequencies, namely $\omega_F = 300, 600$ and $800$ MHz are not statistically significant because the changes in these cases fall within the range of variations observed among different simulation runs with the baseline condition.

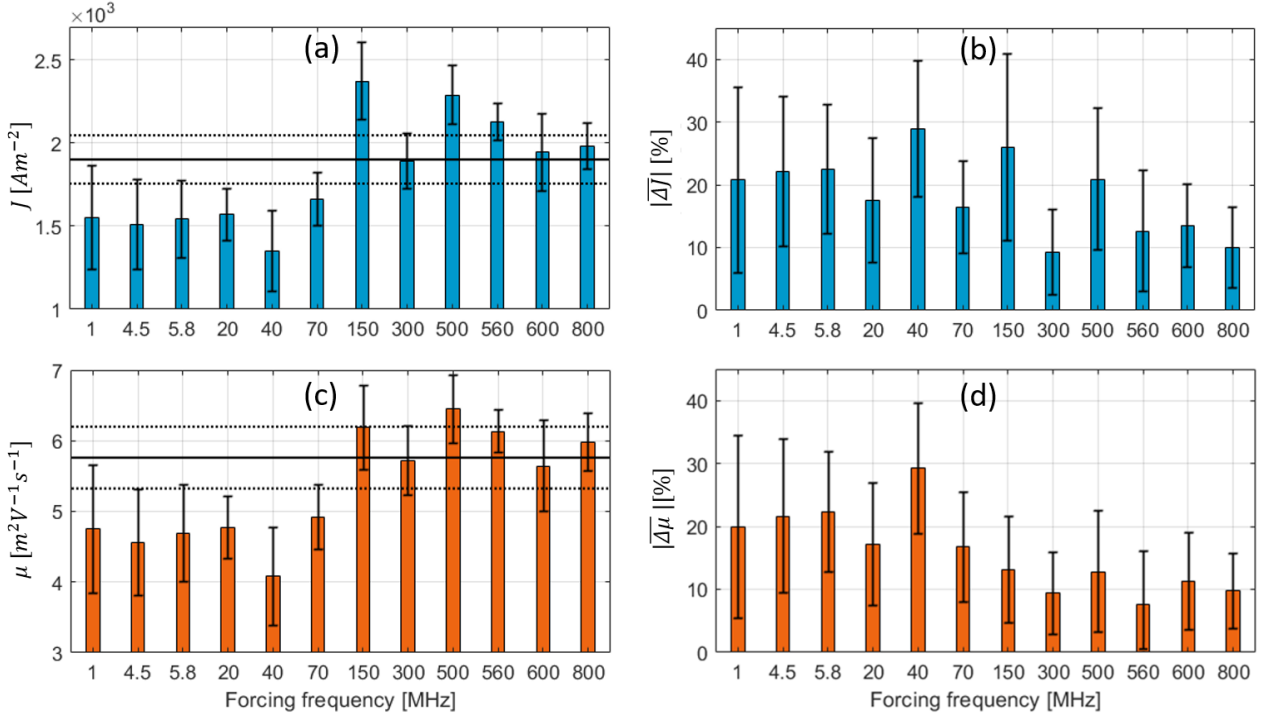

Figure 4: Electrons' axial current density ($J$) and electrons' axial mobility ($\mu$) from the 1D azimuthal simulations with various forcing frequencies. The color bars represent the spatiotemporal mean value (over $4 - 10$ $\mu s$ and entire domain) over 8 simulation repetitions. Plots (a) and (c) display the absolute values of $J$ and $\mu$, and (b) and (d) show the normalized change of $J$ and $\mu$ in the forced simulations with respect to the respective quantities in the baseline case. The error bars indicate one standard deviation among the 8 simulation repetitions in each case. The solid and the dotted black lines, respectively, represent the average and one standard deviation corresponding to 8 simulation repetitions for the baseline case.

**4.2. 2D radial-azimuthal configuration**

In the simulated radial-azimuthal E×B plasma configuration, the plasma dynamics is predominantly influenced by the existence of two instabilities, namely the ECDI and the MTSI. The MTSI is observed to have a radial wavenumber component [35][43], which prevents it from being captured in purely azimuthal simulations. Its presence notably modifies the fluctuations spectra, significantly influencing the coupling of excitation waves to the plasma and the energy cascade process.

Figure 5 and Figure 6 present the variations in the frequency and wavenumber spectra in the presence of various forcing frequencies and amplitudes.

In baseline conditions (without excitation, solid black lines), the wavenumbers of the ECDI and the MTSI modes that develop in the simulation are $7.5 \times 10^3$ and $1.5 \times 10^3$ $m^{-1}$, respectively. These values align well with the theoretical predictions from Eq. 1 and Eq. 2, which yield wavenumbers of $7.0 \times 10^3$ and $1.3 \times 10^3$ $m^{-1}$ for the ECDI and the MTSI, respectively.

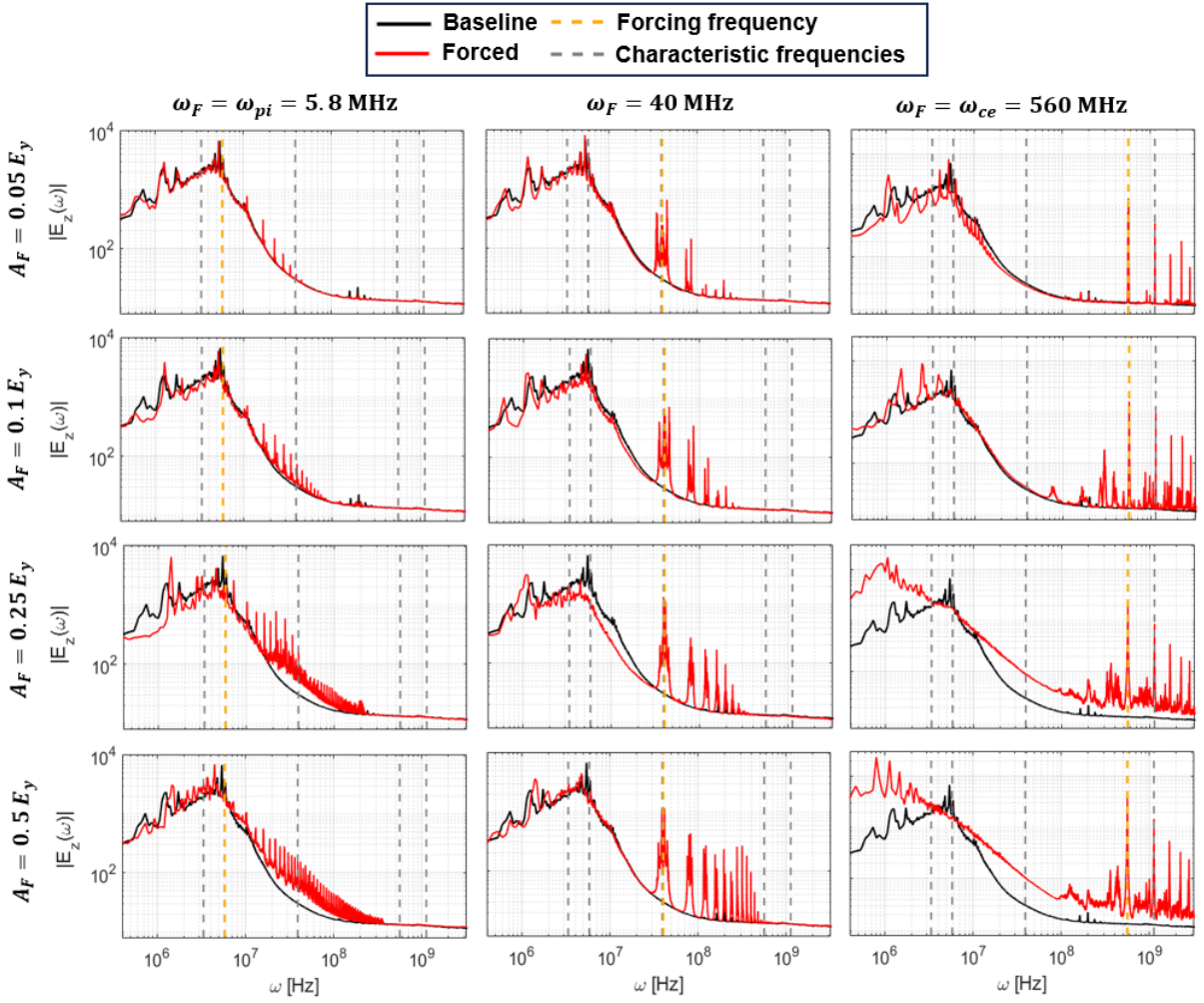

Figure 5: Variation of the frequency spectrum relative to the baseline case from the radial-azimuthal simulations with various forcing frequencies and amplitudes. In each case, the spectrum represents the average of the temporal FFT of azimuthal electric field ($E_z$) signal over all azimuthal and radial positions and over 3 simulation repetitions. The characteristic frequencies displayed in grey dashed lines from left to right correspond to theoretical ion acoustic frequency ($\omega_{IA}$), ion plasma frequency ($\omega_{pi}$), Hall circulation frequency ($\omega_E$) and first and second harmonics of electron cyclotron frequency ($\omega_{ce}, 2\omega_{ce}$).

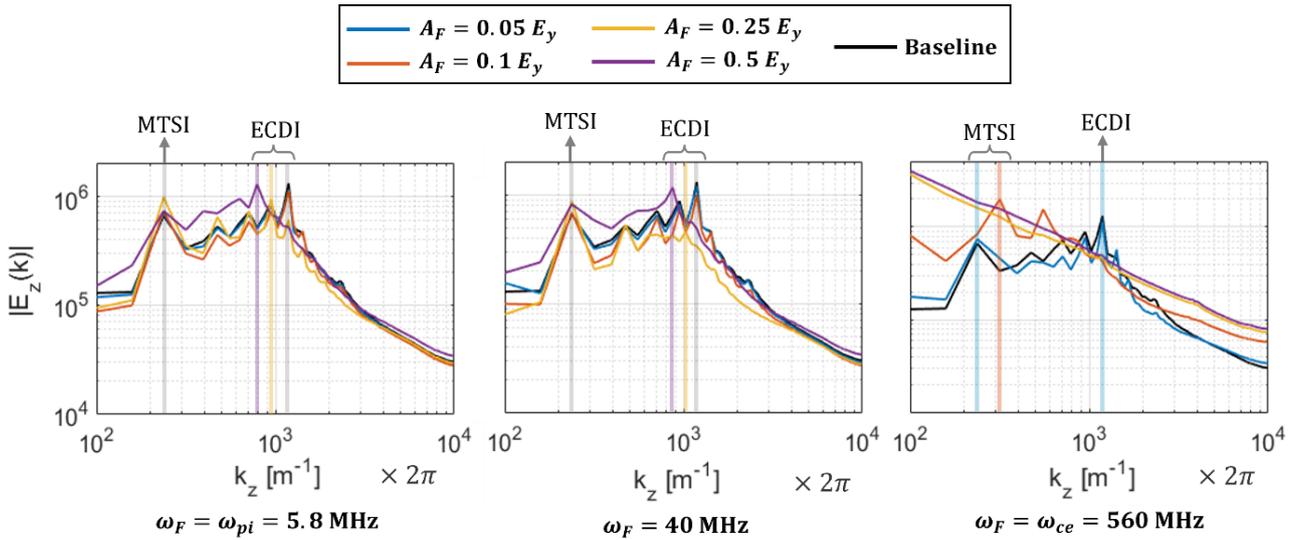

Figure 6: Variation of the azimuthal wavenumber ($k_z$) spectrum relative to the baseline case from the radial-azimuthal simulations with various forcing frequencies and amplitudes. In each case, the spectrum represents the spatiotemporally averaged (over $5 - 15$ $\mu s$ [35-45 $\mu s$ absolute time] and all radial positions) spatial FFT of azimuthal electric field ($E_z$) signal averaged over 3 simulation repetitions.

The different forcing frequencies examined in Figures 5 and Figure 6 (5.8 MHz, 40 MHz, and 560 MHz) exhibit distinct effects on the plasma instabilities' spectra, with varying impacts based on the amplitude of forcing. Forcing at the plasma frequency (5.8 MHz) primarily excites high-frequency oscillations within the 10–200 MHz range, particularly at higher forcing amplitudes ($A_F \geq 0.25 E_y$). At low forcing amplitudes (e.g., $A_F = 0.05 E_y$), the frequency spectrum closely resembles the baseline case, with minor deviations. However, as the forcing amplitude increases, spectral energy spreads to higher frequencies beyond the forcing frequency, activating modes that were less prominent in the baseline. This suggests that 5.8 MHz forcing can drive additional energy into high-frequency oscillations.

Forcing at 40 MHz introduces waves at the forcing frequency itself and excites higher harmonics as the forcing amplitude increases. At an intermediate forcing amplitude, particularly $A_F = 0.25 E_y$, energy is noticeably redistributed, with reduced spectral power in the mid-frequency range (2–20 MHz). This frequency range overlaps with the typical ECDI modes, leading to a significant suppression of this instability.

Forcing at the electron cyclotron frequency (560 MHz) at sufficiently high amplitudes ($A_F \geq 0.25 E_y$) induces substantial changes in the instability spectrum, significantly altering the plasma dynamics. Forcing at 560 MHz at these amplitudes effectively destroys the ECDI's wave structure, as reflected in the reduced spectral power in the ECDI frequency range. Concurrently, high-frequency oscillations are strongly enhanced, with pronounced spectral energy around the forcing frequency and its harmonics. Moreover, the enhanced inverse energy cascade leads to remarkable growth of lower frequencies and larger spatial structures, as observed in both the frequency (Figures 5) and wavenumber spectra (Figure 6). This behavior is indicative of an enhanced coupling across scales, where high-frequency excitation affects the large-scale plasma organization. This suggests that 560 MHz forcing not only disrupts the ECDI but also alters the energy distribution across the entire spectrum. This behavior can be attributed to the fact that excitation at electron cyclotron frequency can resonate with electrons in their cyclotron motion and lead to significant demagnetization of electrons as also observed in the 1D cases with similar excitation frequencies.

In comparing the effects of unsteady forcing in the radial-azimuthal geometry to those in the purely azimuthal configuration, rather similar unsteady forcing effects are observed at forcing frequencies of 5.8 MHz and 40 MHz in both setups. However, the specific impacts differ due to variations in the baseline instability spectrum introduced by the inclusion of the radial dimension. In contrast, forcing at 560 MHz in the radial-azimuthal geometry produces a substantially stronger influence than in the purely azimuthal geometry. This enhanced response is likely to be attributed to the stronger resonant wave-particle interaction at the electron cyclotron frequency which creates more substantial de-magnetization of electrons. This can be inferred from Figure 5, which shows that the excited wave at electron cyclotron frequency has a substantially large amplitude (comparable to ECDI's amplitude in the baseline case). In addition, the presence of the MTSI in the 2D setup facilitates an energy cascade toward low-frequency, long-wavelength modes, likely by expanding energy transfer pathways, further contributing to the notably enhanced effect of excitation at this frequency in the 2D configuration.

Figure 6 shows that at forcing frequencies of 5.8 MHz and 40 MHz, the wavenumber of the ECDI shifts toward lower values, with the extent of this shift increasing as the forcing amplitude rises. In contrast, the wavenumber of the MTSI remains unaffected across these forcing conditions.

Looking at Figure 7, which presents the variations in the electrons' axial current density and mobility with forcing frequency and amplitude, it is observed that forcing at $\omega_F = 40$ MHz and $A_F = 0.25\ E_y$ results in the largest reduction in the electron current. Specifically, this condition reduces the electron current by approximately 38% and decreases electron mobility by around 30 % relative to the baseline condition. This is consistent with the trend seen in purely azimuthal case (Figure 4), where the greatest reduction in the electrons' axial transport – by approximately 30% – also occurred at this forcing frequency.

Forcing at the ion plasma frequency (5.8 MHz) also reduces electrons' current density and mobility by up to approximately 24% at forcing amplitudes of $A_F = 0.1 E_y$ and $0.25\ E_y$, a reduction similar to that observed in the purely azimuthal configuration.

In contrast to both observations above, excitation at the electron cyclotron frequency (560 MHz) with a low amplitude ($A_F = 0.05 E_y$) significantly decreases electron current and mobility by 37% and 24%, respectively. However, as the forcing amplitude increases, the electrons' axial transport is drastically enhanced, rising by more than two orders of magnitude at $A_F = 0.25 E_y$ and $0.5\ E_y$. This level of enhancement in the electrons' transport is likely due to increased turbulence and their direct contribution to transport as well as heating of particles which leads to a breakdown of magnetic confinement of electrons due to elevated electron temperatures (substantially elongated electrons' cyclotron orbit along the axial direction) under these forcing conditions.

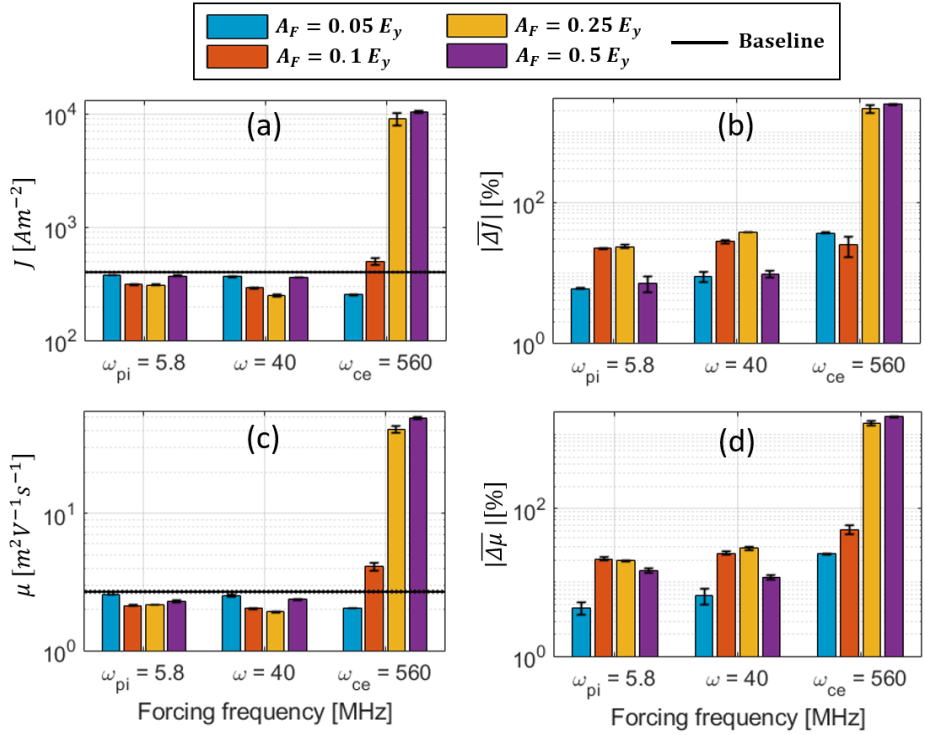

Figure 7: Electrons' axial current density ($J$) and electrons' axial mobility ($\mu$) from the radial-azimuthal simulations with various forcing frequencies and amplitudes. The color bars represent the spatiotemporal mean value (over $5-15\ \mu s$ [35-45 $\mu s$ absolute time] and entire domain) averaged over 3 simulation repetitions. Plots (a) and (c) display the absolute values of $J$ and $\mu$, and (b) and (d) show the normalized change of $J$ and $\mu$ in the forced simulations with respect to the respective quantities in the baseline case. The error bars indicate one standard deviation among the 3 simulation repetitions in each case. The solid and the dotted black lines represent the average and one standard deviation corresponding to 3 simulation repetitions for the baseline case.

For further insight, the overall impacts of the unsteady forcing and, consequently, the altered effect of the instabilities on the particles' velocity distribution functions, as well as on the plasma profiles, are provided in Appendix A. The plots in Appendix A show that excitation at 560 MHz with $A_F \geq 0.1 E_y$ results in notable heating of the electrons and the ions and substantial broadening of their velocity distributions along both radial and azimuthal directions.

As an additional study, we have provided in Appendix B an analysis of unsteady forcing along the radial direction. The results indicate that the radial forcing (with a radially oscillating electric field) generally has minimal impact on the spectral amplitudes and on the electron transport, with the exception of high-amplitude excitation at the electron cyclotron frequency ($\omega_F$ =560 MHz, $A_F \geq 0.5 E_y$), which reduces electrons' axial current by 5-8 %.

**Section 5: Closer examination of the observations**

In the previous section, we identified that an excitation frequency of $\omega_F = 40$ MHz with the amplitude of $A_F = 0.25 E_y$ is the most effective to reduce electron transport among the studied excitation scenarios. Building on this finding, we now investigate how fine variations around this specific frequency and amplitude can further impact the instabilities' spectra and the consequent electron transport. This analysis serves as a sensitivity study for fine-tuning frequency and amplitude, aiming to determine whether minor adjustments might lead to an even greater reduction in electron transport, which is an essential step in optimizing these parameters for practical applications. To achieve this, the forcing frequency is varied from 10 MHz to 100 MHz in 10 MHz increments, while maintaining a constant amplitude of $0.25\ E_y$. Similarly, the forcing amplitude is adjusted from $0.05\ E_y$ to $0.5\ E_y$ incrementally, with the frequency fixed at 40 MHz.

The variations in the frequency and wavenumber spectra with the forcing frequency and amplitude are presented in Figure 8 and Figure 9. Across all parameter studies, the peak associated with the ECDI is consistently weakened.

Notably though, a forcing frequency around 40 MHz combined with an amplitude near $0.25E_y$ still represents the most substantial impact on reducing the ECDI amplitude. This reduction in ECDI amplitude directly contributes to a significant decrease in electron transport, highlighting the effectiveness of this specific frequency-amplitude configuration in controlling this instability mode and enhancing transport suppression.

Furthermore, the average electrons' axial current and mobility plots in Figure 10 and Figure 11 suggests that excitation at 40 MHz with amplitude $0.1E_y - 0.3E_y$ represent optimal range of configuration for the minimization of electrons' cross-field transport in the studied 2D plasma configuration.

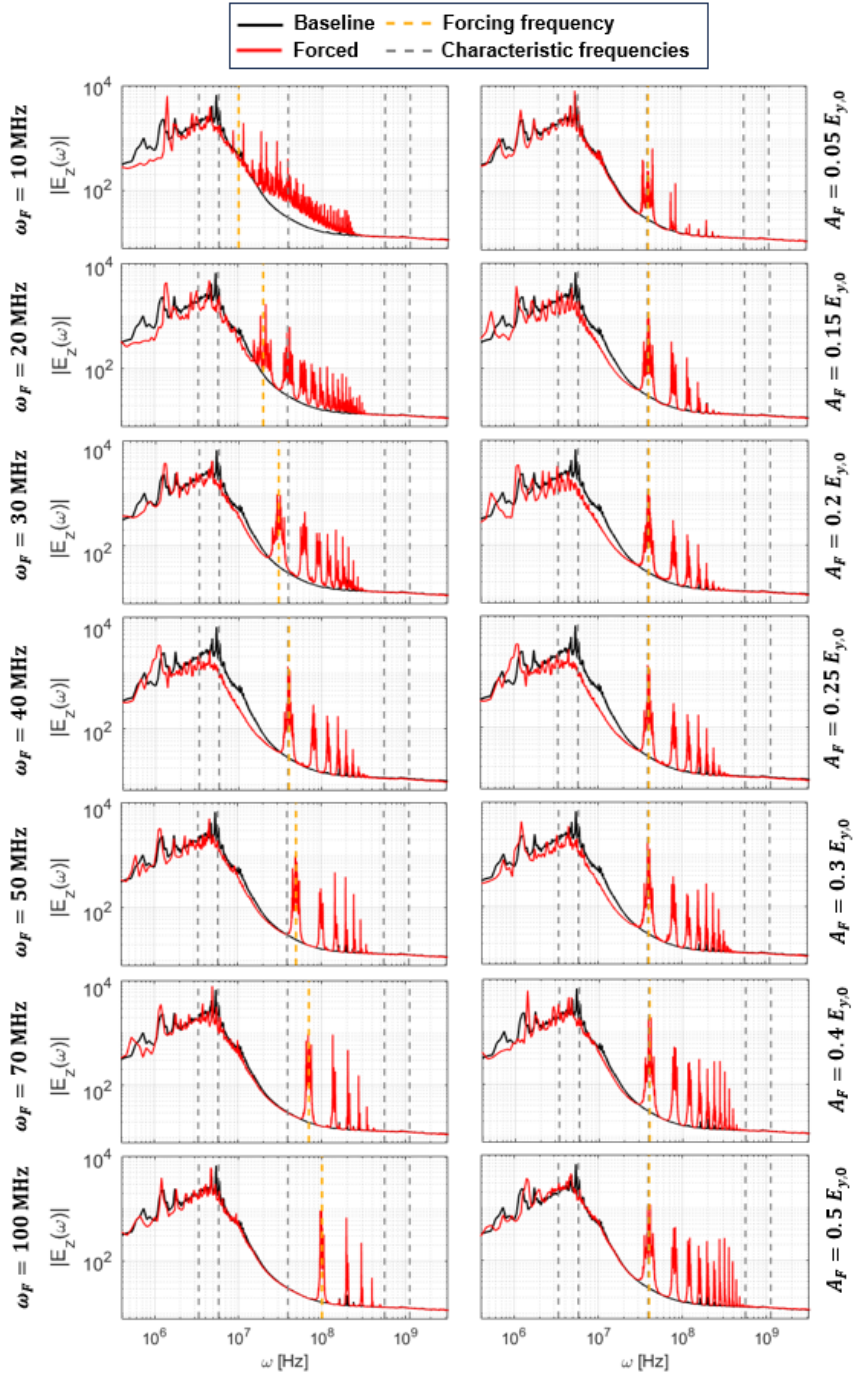

Figure 8: Variation of the frequency spectrum relative to the baseline case from the radial-azimuthal simulations with various forcing frequencies ($10 - 100\ MHz$) and amplitudes ($0.05 - 0.5\ E_y$). In the left-column plots, the forcing amplitude is fixed at $0.25E_y$, and in the right-column plots the forcing frequency is fixed at 40 MHz. In each case, the spectrum represents the average of the temporal FFT of azimuthal electric field ($E_z$) signal over all azimuthal and radial positions and over 3 simulation repetitions. The characteristic frequencies displayed in grey dashed lines from left to right correspond to theoretical ion acoustic frequency ($\omega_{IA}$), ion plasma frequency ($\omega_{pi}$), Hall circulation frequency ($\omega_E$) and first and second harmonics of electron cyclotron frequency ($\omega_{ce}, 2\omega_{ce}$).

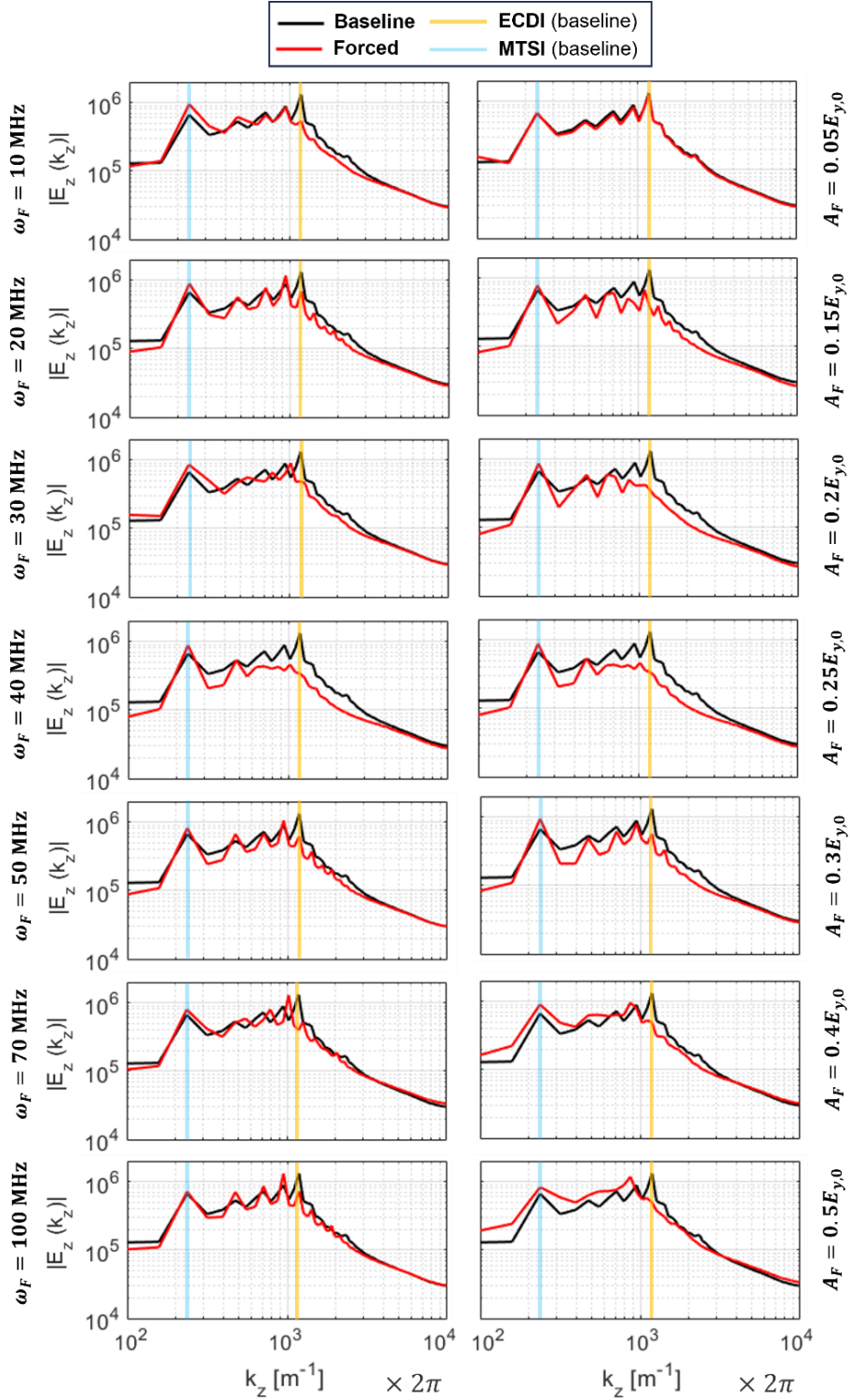

Figure 9: Variation of the azimuthal wavenumber ($k_z$) spectrum relative to the baseline case from the radial-azimuthal simulations with various forcing frequencies ($10 - 100\ MHz$) and amplitudes ($0.05 - 0.5\ E_y$). In the left-column plots, the forcing amplitude is fixed at $0.25E_y$, and in the right-column plots the forcing frequency is fixed at 40 MHz. In each case, the spectrum represents the spatiotemporally averaged (over $5 - 15\ \mu s$ [35-45 $\mu s$ absolute time] and all radial positions) spatial FFT of azimuthal electric field ($E_z$) signal averaged over 3 simulation repetitions. The yellow and blue lines indicate the theoretical $k_z$ of the ECDI and the MTSI based on Eq. 1 and Eq. 2, respectively.

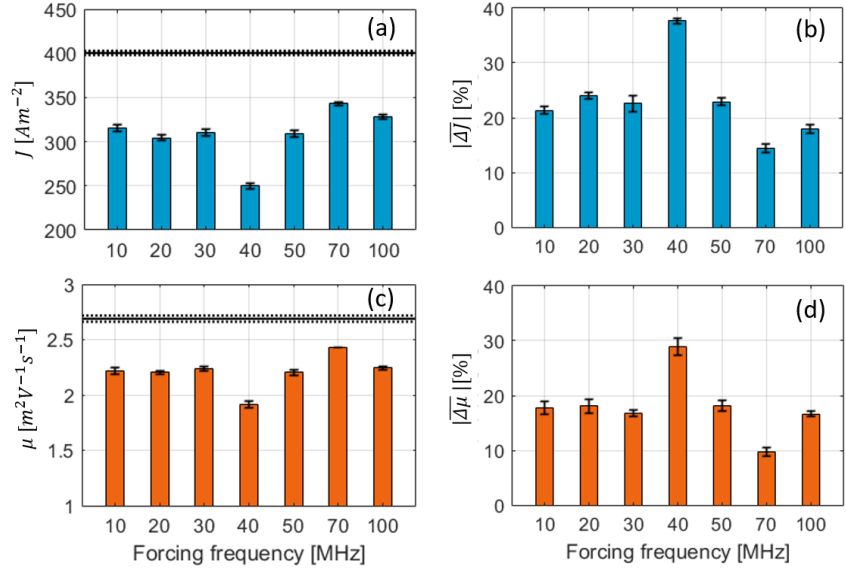

Figure 10: Electrons' axial current density ($J$) and electrons' axial mobility ($\mu$) from the radial-azimuthal simulations with various forcing frequencies ($10 - 100\ MHz$). The forcing amplitude is fixed at $0.25 E_y$. The color bars represent the spatiotemporal mean value (over $5 - 15\ \mu s$ [35-45 $\mu s$ absolute time] and entire domain) averaged over 3 simulation repetitions. Plots (a) and (c) display the absolute values of $J$ and $\mu$, and (b) and (d) show the normalized change of $J$ and $\mu$ in the forced simulations with respect to the respective quantities in the baseline case. The error bars indicate one standard deviation among the 3 simulation repetitions in each case. The solid and the dotted black lines represent the average and one standard deviation corresponding to 3 simulation repetitions for the baseline case.

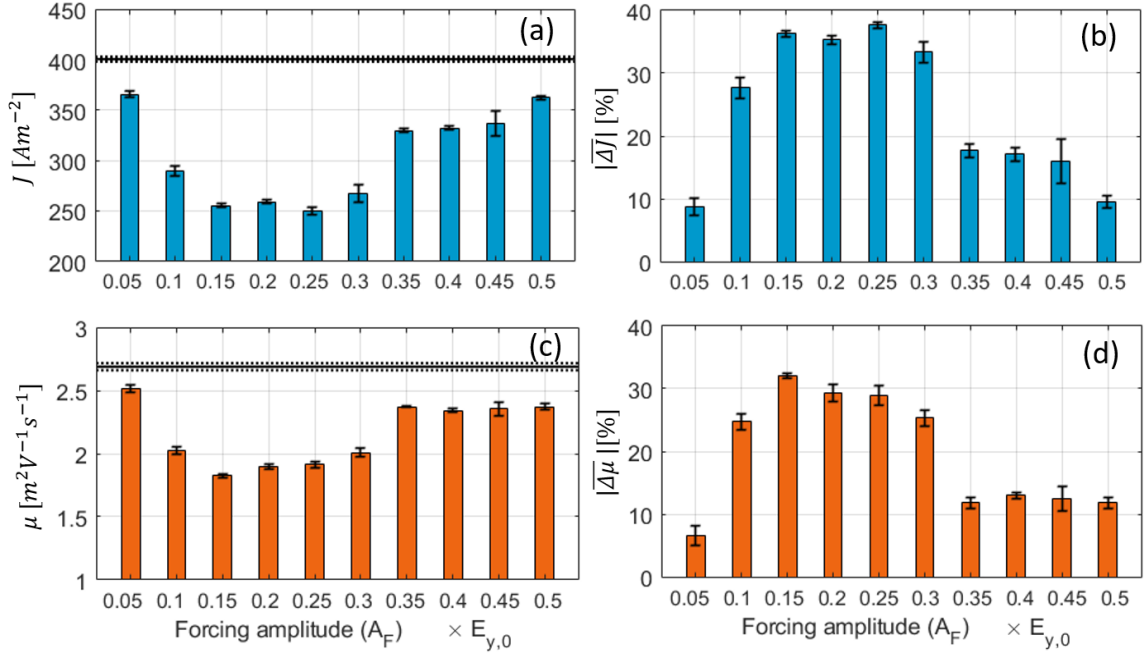

Figure 11: Electrons' axial current density ($J$) and electrons' axial mobility ($\mu$) from the radial-azimuthal simulations with various forcing amplitude ($0.05 - 0.5\ E_y$). The forcing frequency is fixed at 40 MHz. The color bars represent the spatiotemporal mean value (over $5 - 15\ \mu s$ [35-45 $\mu s$ absolute time] and entire domain) averaged over 3 simulation repetitions. Plots (a) and (c) display the absolute values of $J$ and $\mu$, and (b) and (d) show the normalized change of $J$ and $\mu$ in the forced simulations with respect to the respective quantities in the baseline case. The error bars indicate one standard deviation among the 3 simulation repetitions in each case. The solid and the dotted black lines represent the average and one standard deviation corresponding to 3 simulation repetitions for the baseline case.

A possible theory for the change in the electron transport due to forcing as we reported in this paper could be offered as follows: the shifting in the phase of the azimuthal electric field oscillations $\tilde{E}_z$ and the electron density oscillations $\tilde{n}_e$ because of the redistribution of the frequency spectra of the contributing instabilities, which occurs through either frequency lock-in or variation in the energy cascade behavior. If the cross-correlation term between the two fields $<\tilde{n}_e \tilde{E}_z>$ is changed as a result, this will affect the resulting electron transport.

Confirming frequency locking in a simulation environment would require spectral phase analysis or coherence diagnostics, such as bicoherence analysis [44], to assess the persistence of phase relationships between natural oscillations and the applied forcing. More generally, techniques that assess triad interactions [45], based on frequency and wavevector matching, can reveal the presence and dynamics of energy transfer between natural oscillations and the applied forcing. These investigations and analyses have been left for future work.

From an experimental perspective, with an outlook toward validation of the observed behaviors in this work in real systems, similar approaches as above may be employed. Bicoherence analysis, as well as wavelet-based techniques, along with the employment of time-resolved diagnostics, such as high-speed probe measurements of fluctuating fields, can be used to detect signatures of phase synchronization between natural fluctuation waves in the plasma and the external modulator, often referred to as entrainment.

**Section 6: Conclusion**

This study has demonstrated that externally applied unsteady electrostatic forcing can substantially influence instability-driven electron transport in E×B plasma discharges. Through one- and two-dimensional kinetic simulations, we explored how the frequency and amplitude of an imposed axial electric field affect the spectral energy distribution of plasma instabilities and the resulting cross-field transport. The results suggest that nonlinear interactions, such as possible energy pathway modulation and frequency locking effects, appear to govern the system's response to excitation, offering routes to either suppress or enhance instability-induced electron mobility.

In the 1D azimuthal simulations, forcing at frequencies below 150 MHz, particularly near 40 MHz, consistently suppressed the ECDI, reducing its spectral amplitude and shifting energy toward longer wavelengths and lower frequencies. This spectral redistribution correlated with a notable decrease in axial electron transport, with the strongest reduction – up to 30 % – observed at 40 MHz. In contrast, forcing near or above the electron cyclotron frequency intensified high-frequency oscillations and led to a recovery or increase in transport, attributed to electron demagnetization and enhanced energy spread.

The 2D radial-azimuthal simulations exhibited similar trends at corresponding forcing frequencies, validating the key mechanisms observed in 1D. However, the inclusion of radial dynamics and the presence of the MTSI introduced additional coupling between scales. The MTSI, with its radial wavenumber component, was observed to enable stronger inverse energy cascades under high-frequency forcing. As a result, excitation at 560 MHz produced even more pronounced spectral broadening and stronger transport enhancement at high amplitudes, reaching increases of more than two orders of magnitude in axial current. At moderate amplitudes, however, the same excitation could suppress transport, highlighting the non-monotonic and amplitude-sensitive nature of the response.

Together, the above results establish a clear link between external excitation parameters, instability spectra, and the resulting transport behavior in E×B plasmas. The observed sensitivity to frequency and amplitude suggests that precise tailoring of forcing conditions can be used to suppress transport-enhancing instability modes while preserving or enhancing confinement. Importantly, reductions in electron transport were achieved through modest excitation amplitudes, indicating the potential for practical implementations with limited power overhead.

It is also noteworthy that the extensive simulation studies in this work were made computationally feasible partly through the use of reduced-order particle-in-cell method [41], which retains kinetic fidelity while offering substantial speed-up over conventional PIC schemes. This enabled the broad parametric exploration in 2D configuration across frequency-amplitude space reported in the paper.

Our findings from this work may offer new insights into how spectral redistribution and nonlinear coupling govern instability-induced transport in E×B plasmas. The observed frequency-dependent responses highlight the complexity of energy exchange processes between instability modes under external excitation. Experimental validation of these observations represents a key next step, particularly in assessing the feasibility of applying targeted forcing in real plasma devices and evaluating the degree of correspondence between simulation predictions and experimental observations. Additionally, further analysis of energy transfer pathways, such as the conditions that favor inverse cascades or frequency locking, will be essential to deepen physical understanding of the mechanisms responsible for the observed transport modifications.

**Acknowledgments**:

There was no funding supporting this research.

MR and FF gratefully acknowledge the computational resources and support provided by the Imperial College Research Computing Service (http://doi.org/10.14469/hpc/2232).

**Appendix: Supplementary results**

**A. Variations in plasma profiles and particles' distribution functions in the radial-azimuthal configuration**

The time-averaged radial profiles of plasma properties in the radial-azimuthal plasma configuration under different excitation conditions are presented in Figure 12. Furthermore, the time-averaged particles' velocity distribution functions along radial and azimuthal directions from the corresponding simulations are provided in Figure 13. These figures illustrate the influences of excitations with various frequencies and amplitudes and altered instabilities spectra on the plasma properties and particles velocity spreads.

In particular, in the case of excitation at 560 MHz with $A_F \geq 0.1 E_y$, the large-scale structures developed due to the intensified inverse energy cascade lead to remarkable heating of both electrons and ions in the radial and azimuthal directions. This effect is evident in the temperature profiles and is manifested as the pronounced broadening of the velocity distribution functions.

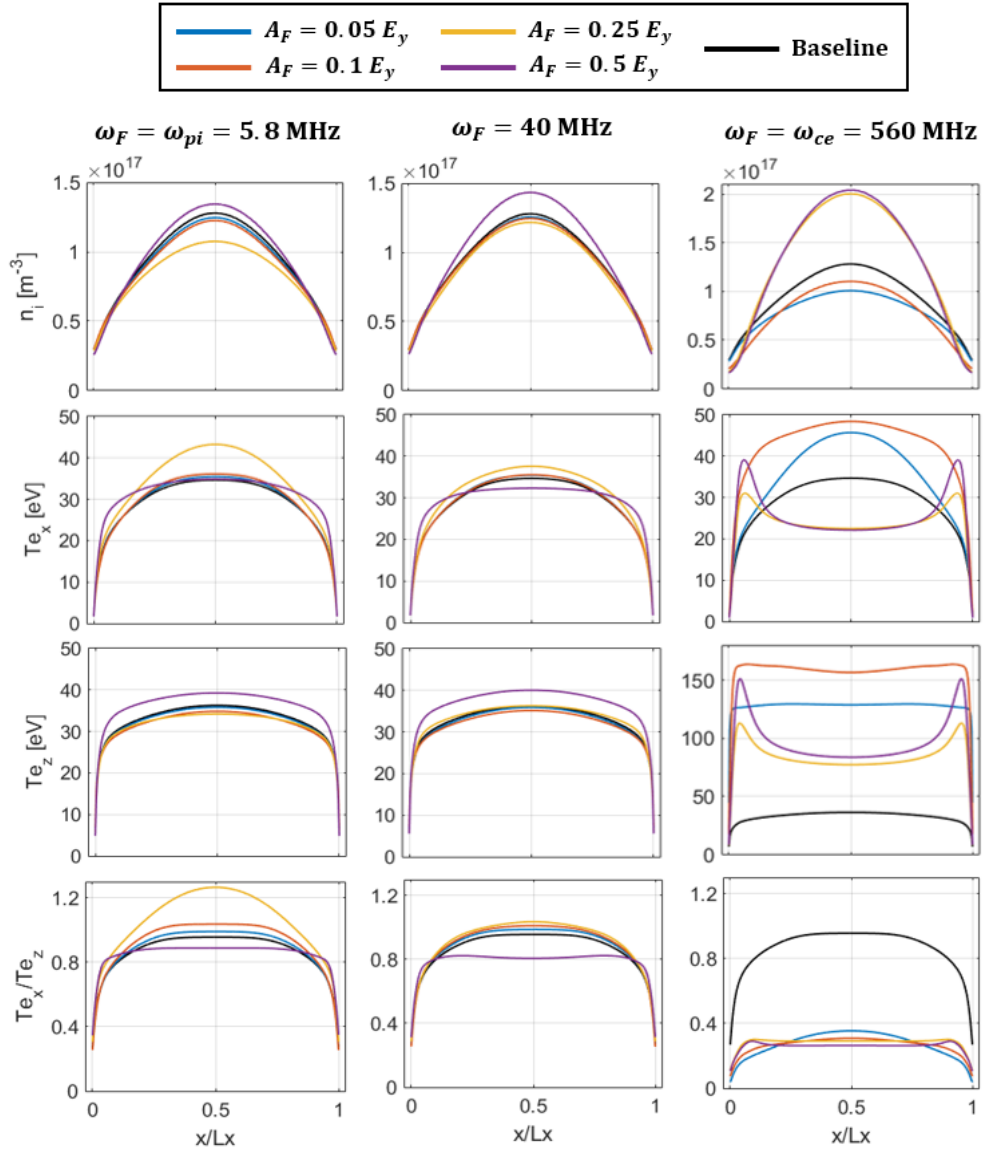

Figure 12: Time-averaged (over $5-15\ \mu s$ [35-45 $\mu s$ absolute time]) radial profiles of the plasma properties from the radial-azimuthal simulations with various forcing frequencies and amplitudes averaged over all azimuthal locations and over 3 simulation repetitions. The rows from top to bottom represent ion number density ($n_i$), radial electron temperature ($T_{ex}$), azimuthal electron temperature ($T_{ez}$), and the ratio of the radial to azimuthal electron temperature ($T_{ex}/T_{ez}$). Note that in plots of $n_i$ and $T_{ez}$ for $\omega_F = 560\ MHz$, the scale of $y$-axis is different than the corresponding plots for the other forcing frequencies.

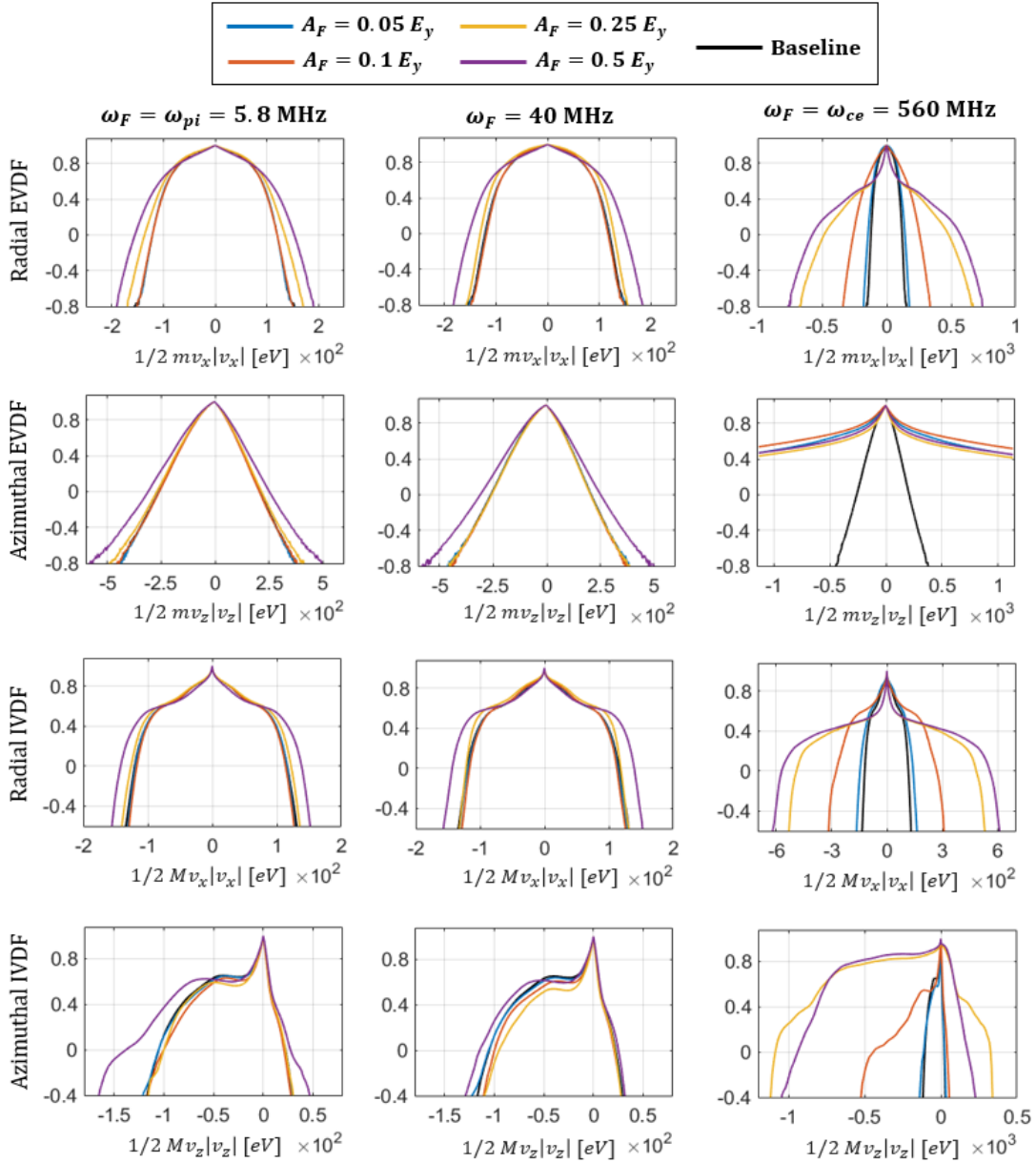

Figure 13: Time-averaged (over $5-15\ \mu s$ [35-45 $\mu s$ absolute time]) electrons' and ions' velocity distribution functions (EVDF and IVDF, respectively) along the radial and azimuthal directions from the radial-azimuthal simulations with various forcing frequencies and amplitudes. The distribution functions correspond to the electrons or ions within the entire domain and are averaged over 3 simulation repetitions. Note that in the plots for $\omega_F = 560\ MHz$, the scale of $x$-axis is different than the corresponding plots for the other forcing frequencies.

**B. Forcing in the radial direction**

In this section, the results of the investigation of unsteady forcing along the radial direction are presented. For this purpose, an oscillating electric field $E_x = A_F \sin(2\pi \omega_F t)$ is applied along the radial direction, using the same frequencies and amplitudes as those explored in Section 4.2. The resulting instabilities spectra in the presence of the radial forcings are compared with those under axial forcings in Figure 14. It is seen that the impact of radial forcing on the spectral amplitudes of instabilities is minimal. Forcing at 40 MHz and 560 MHz excites waves at the respective forcing frequency and their harmonics; however, it has no appreciable effect on the energy of the other frequencies in the spectrum.

These minor variations in the spectral amplitudes of instability translate into minimal impact on the electrons axial transport, as shown in Figure 15. The changes in electron current density and mobility remain below 3%, except when excitation occurs at the electron cyclotron frequency with high-amplitude. In this specific case ($\omega_F =560$ MHz, $A_F \geq 0.5 E_y$), unlike with axial forcing, where electron transport increased by two orders of magnitude, high-amplitude radial forcing leads to a reduction in the axial electron current by 5-8%.

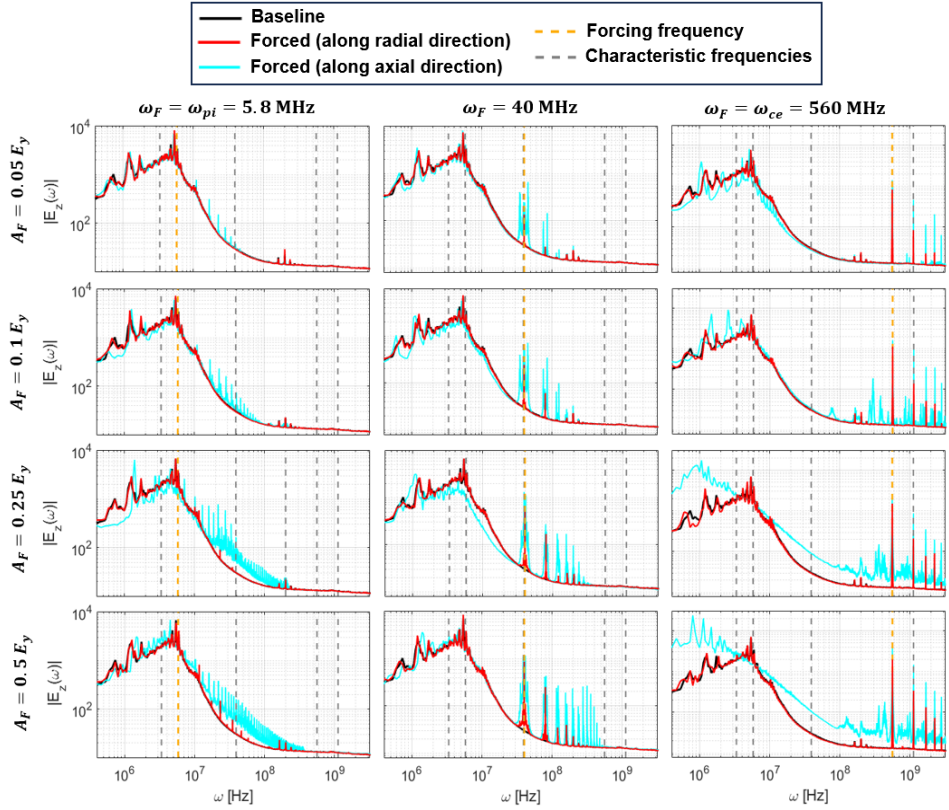

Figure 14: Frequency spectra from the radial-azimuthal simulations with forcing along radial direction compared to the baseline and the forced simulations along the axial direction. In each case, the spectrum represents the average of the temporal FFT of azimuthal electric field ($E_z$) signal over all azimuthal and radial positions and over 3 simulation repetitions. The characteristic frequencies displayed in grey dashed lines from left to right correspond to theoretical ion acoustic frequency ($\omega_{IA}$), ion plasma frequency ($\omega_{pi}$), Hall circulation frequency ($\omega_E$) and first and second harmonics of electron cyclotron frequency ($\omega_{ce}, 2\omega_{ce}$).

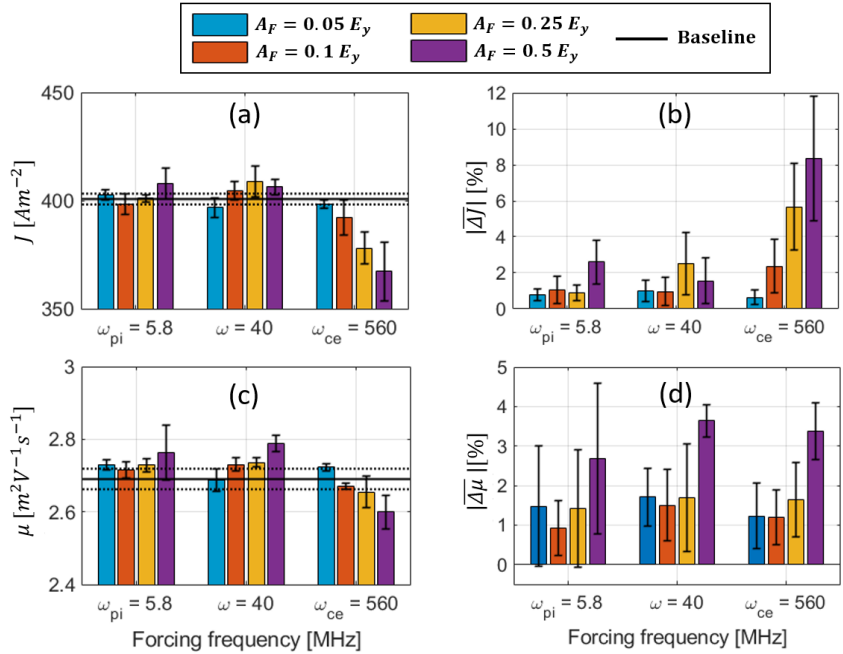

Figure 15: Electrons' axial current density ($J$) and electrons' axial mobility ($\mu$) from the radial-azimuthal simulations with forcing along radial direction. The color bars represent the spatiotemporal mean value (over $5-15\ \mu s$ [35-45 $\mu s$ absolute time] and entire domain) value averaged over 3 simulation repetitions. Plots (a) and (b) display the absolute values of $J$ and $\mu$, and (b) and (d) show the normalized change of $J$ and $\mu$ in the forced simulations with respect to the respective quantities in the baseline case. The error bars indicate one standard deviation among the 3 simulation repetitions in each case. The solid and the dotted black lines represent the average and one standard deviation corresponding to 3 simulation repetitions for the baseline case.

**Data Availability Statement**:

The simulation data that support the findings of this study are available from the corresponding author upon reasonable request.